\documentclass[conference]{IEEEtran}
\usepackage{amssymb, amsmath}
\usepackage{graphicx}
\usepackage{color}
\usepackage{flexisym}
\usepackage{breqn}

\usepackage{cite}

\ifCLASSINFOpdf
\else
\fi
\begin{document}
\title{Performance Analysis of a Scalable DC Microgrid Offering Solar Power Based Energy Access and Efficient Control for Domestic Loads}

\author{\IEEEauthorblockN{Abu~Shahir~Md.~Khalid~Hasan}
\IEEEauthorblockA{Department
of Electrical \\and Electronic Engineering\\
Daffodil International University\\ Dhaka, Bangladesh.\\
Email: khalid.eee@diu.edu.bd}
\and
\IEEEauthorblockN{Dhiman~Chowdhury}
\IEEEauthorblockA{Department of Electrical Engineering\\ University of South Carolina\\ United States of America\\
Email: dhiman@email.sc.edu}
\and
\IEEEauthorblockN{Mohammad~Ziaur~Rahman~Khan}
\IEEEauthorblockA{Department
of Electrical \\and Electronic Engineering\\
Bangladesh University of\\
Engineering and Technology\\ Dhaka, Bangladesh.\\
Email: zrkhan@eee.buet.ac.bd}}

\maketitle

\begin{abstract}
DC microgrids conform to distributed control of renewable energy sources which ratifies efficacious instantaneous power sharing and sustenance of energy access among different domestic Power Management Units (PMUs) along with maintaining stability of the grid voltage. In this paper design metrics and performance evaluation of a scalable DC microgrid are documented where a look-up table of generated power of a source converter complies with the distribution of efficient power sharing phenomenon among a set of two home PMUs. The source converter is connected with a Photovoltaic panel of 300 W and uses Perturb and Observation (P\&O) method for executing Maximum Power Point Tracking (MPPT). A boost average DC-DC converter topology is used to enhance the voltage level of the source converter before transmission. The load converter consists of two parallely connected PMUs each of which is constructed with high switching frequency based Full Bridge (FB) converter to charge an integrated Energy Storage System (ESS). In this paper the overall system is modeled and simulated on MATLAB/Simulink platform with ESSs in the form of Lead Acid batteries connected to the load side of the FB converter circuits and these batteries yield to support marginalized power utilities. The behaviour of the system is tested in different solar insolation levels along with several battery charging levels of 12 V and 36 V to assess the power efficiency. In each testbed the efficiency is found to be more than 93\% which affirm the reliability of the framework and a look-up table is generated comprising the grid and load quantities for effective control of power transmission.
\end{abstract}



%
\IEEEpeerreviewmaketitle

\section{Introduction}
\IEEEPARstart{E}{nergy} poverty is a common problem in many areas around the world. Fossil fuel has deleterious effects on the environment and a major concern for the current situation of global warming. A viable alternative to provide clean energy to the remote places could be the DC microgrid system.
Number of people lacking electricity-access is dwindling in Latin America, North Africa, Middle East, South Asia, China and East Asia but it is still on the increase in  some places like Sub-Saharian Africa. This population will remain unchanged through 2030 based on International Energy Agency (IEA) projection\cite{IEA_2011}. The price of PV components is also falling over the last few years. In addition, more compact and modular design of PV has made the system more acceptable for remote and rural regions. Decentralized electricity generation is imperative because extension of the grid in the remote regions demands extravagant connection cost \cite{Alstone_2015}.
Solar lantern has been used to harness solar energy for lighting and battery charging while solar home system has addressed the electricity need of an individual household in the rural and remote regions \cite{Komatsu_2011}. But solar lantern and solar home system cannot take advantage of multiplexing power generation or storage among several interconnected households. Microgrid brings into contact multiple households and maneuvers one or more generation sources. Microgrid can be a part of larger microgrid or run on recluse islanded-mode. While solar home system needs to be designed for peak load of the household, microgrid can be contrived for the peak demand of the community which will be more cost efficient\cite{Bardouille_2012}. A diesel generator can be used  as another generation source when  the irradiation level is unsatisfactory. Grid connectivity  can be considered as another source of generation\cite{Boroyevich_2010}. The fact that microgrid can manipulate power over the interconnected households infuses confidence for its usage. DC microgrid  uses distributed Point-Of-Load (POL) converters which is much more efficient than always-on central inverters because plethora of DC appliances are utilized in the remote and rural regions. In AC microgrid, inverters experience grid-wide brownouts when correlated peaks are observed on the load\cite{Quetchenbach_2013}. DC microgrid with local storage provides energy individulally to each house hold negating  this effect. Again, the overall efficiency suffers in case of low demand in AC microgrids; whereas the distributed storage system provides better resemblance in case of average demand for the power converters.
The DC-AC conversion losses in the inverters and the central battery bank is more than the DC-DC conversion and distributed battery system used in the DC microgrid \cite{Madduri_2015}-\cite{Brent_2010}.

In this paper, a DC microgrid is simulated where a boost average DC-DC converter is guided by MPPT and then FB converter is used to bring down the voltage to battery level. The system is monitored for distinctive solar insolation levels along with separable load ratings and found to work at similar efficiency. The efficiency is also found to be stable in various loading conditions of the PMUs and different switching frequencies of  FB converter. The overall scalable microgrid architecture is evaluated in regard to power sharing in a set of two PMUs. Each PMU supports a battery intended to transmit consistent power to attached loads like home appliances. This paper concerns with equal and different power sharing between two PMUs to assess power efficiency and reliable power transmission from the source converter to the load side. In the follow-up, an adaptive look-up table has been generated comprising the evaluated power quantities and performance parameters for different switching frequencies and battery voltage levels. The proposed microgrid architecture, PMU circuit configuration and its operation, simulated profiles of microgrid's operation, generated look-up table and a brief explanation of the overall implemented system are articulated in this paper section-by-section.

\begin{figure}[ht!]
\centering
    \includegraphics[width = 3 in, height = 1.5 in] {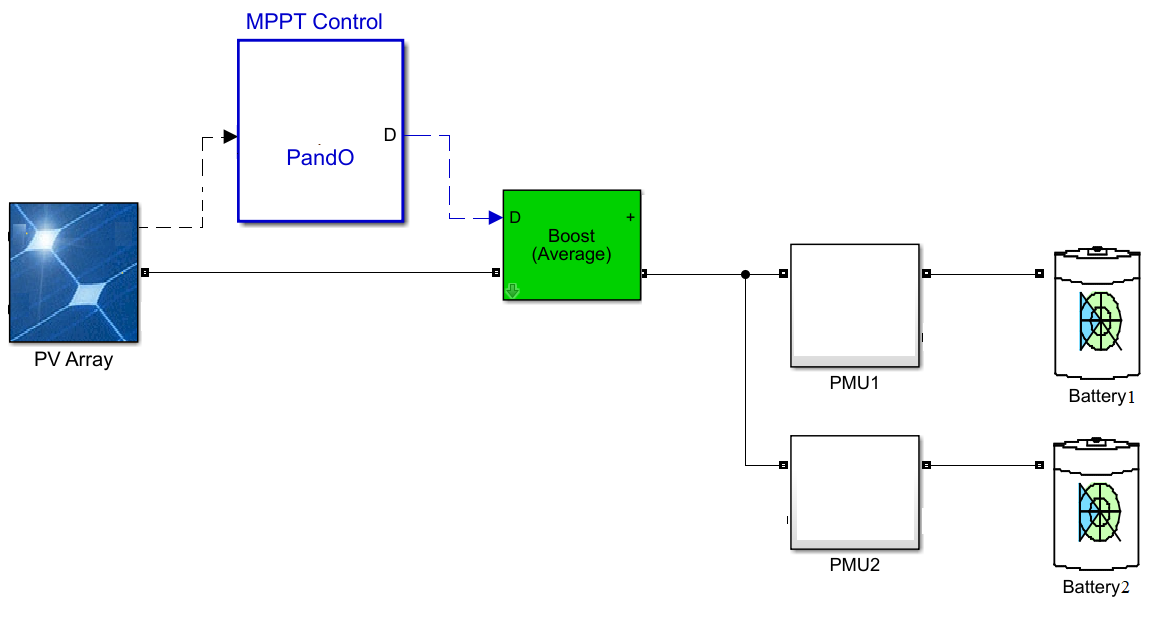}
    \caption{Proposed DC microgrid architecture in MATLAB/ Simulink}
    \label{Block Diagram of DC microgrid}
\end{figure}
\begin{figure}[ht!]
\centering
    \includegraphics[width = 3.5 in, height = 1.3 in] {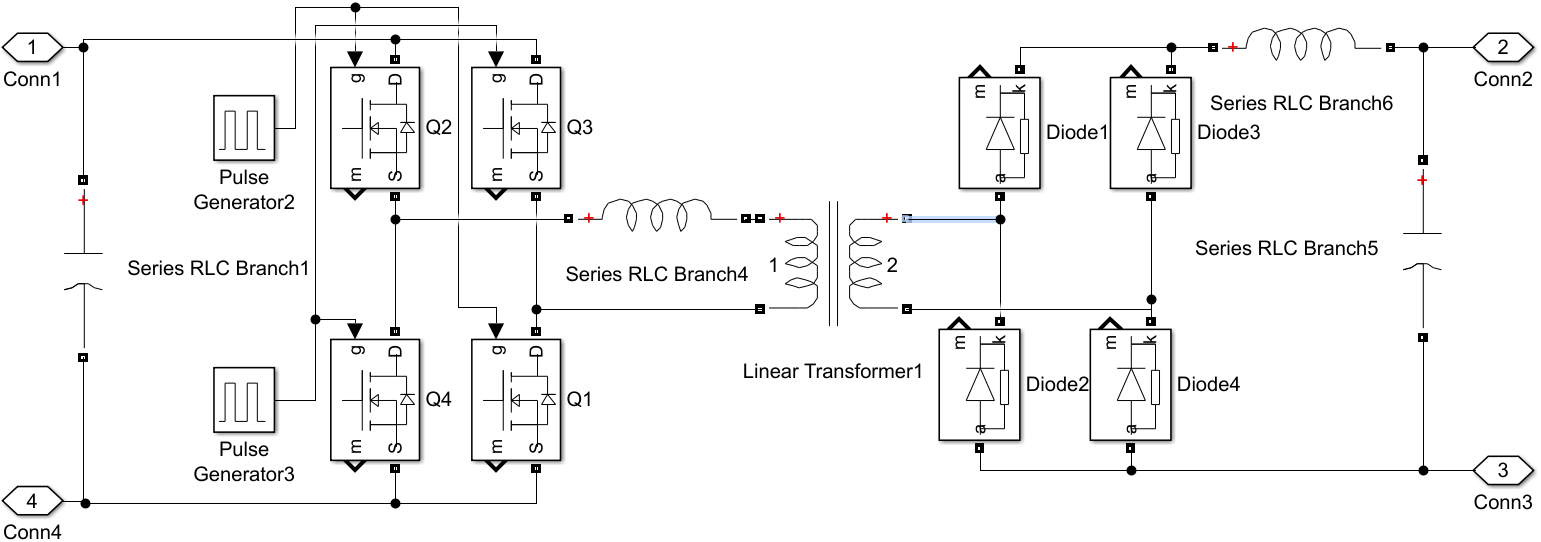}
    \caption{PMU circuit structure in MATLAB/ Simulink}
\end{figure}

\section{Proposed DC Microgrid Architecture}

The PV array acts as a constant current source. Based on irradiation level, the perturb and observation method tracks the maximum power point and generates the duty cycle required for the boost average DC-DC converter to level up the voltage. The PMU consists of a FB converter where MOSFETs are used as the switching devices. The main purpose is to monitor the equal and unequal load sharing of the PMUs and then generate an adaptive look up table for which the power source can be utilized accordingly.

\subsection{Perturb and Observation Algorithm based MPPT Control}
Maximum Power Point Tracking (MPPT) is a methodology used in photovoltaic (PV) converters to continuously adjust the impedance seen by the solar array to keep the PV system operating at, or close to, the peak power point of the PV panel under varying conditions, like changing solar irradiance, temperature, and load. 

The algorithm senses the voltage $V(k)$ and current $I(k)$ to measure the instantaneous power $P(k)$ at the $k^{th}$ sample. Then it compares it with previous power sample $P(k-1)$. If the two power samples are different then it then the voltage sample $V(k)$ is compared with its previous sample $V(k-1)$ and the duty cycle $D_b$ of the boost converter is adjusted accordingly to ensure that the system operates at “maximum power point” (or peak voltage) on the power voltage curve, as shown below. The algorithms account for factors such as variable irradiance (sunlight) and temperature to ensure that the PV system generates maximum power at all times.

\begin{figure}[ht!]
\centering
    \includegraphics[width =1.5 in, height = 0.7 in] {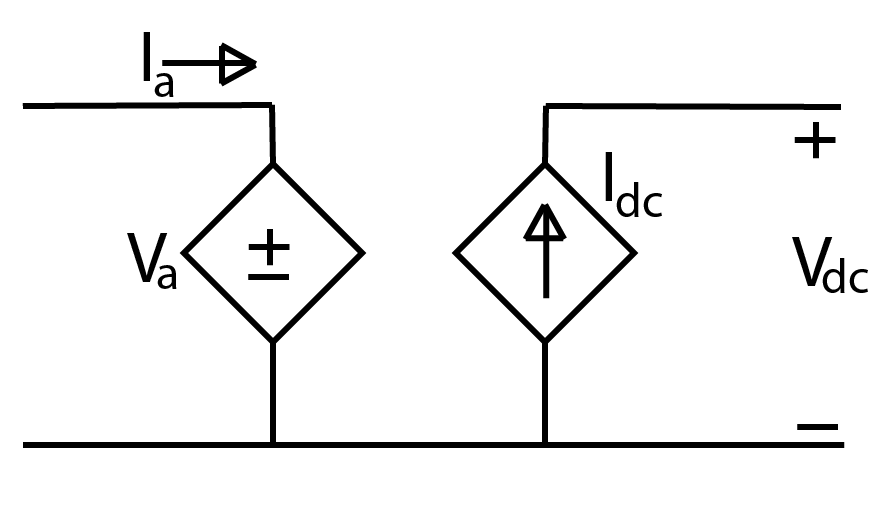}
    \caption{Boost converter average model}
    \label{Average Model for Boost Converter}
\end{figure}

\subsection{Boost Converter Average Model}
The average model for boost converter is designed with a controlled voltage source in the input and a controlled current source in the output. In Fig. \ref{Average Model for Boost Converter}, $V_a$ and $I_a$ are the input current and voltage whereas $V_{dc}$ and $I_{dc}$ are the output current and voltage of the boost converter. The equations for $V_a$ and $I_{dc}$ are
\begin{equation}\label{eq2-}
   V_a(k)=(1-D_b)V_{dc}(k-1)
\end{equation}

\begin{equation}\label{eq1-}
   I_{dc}(k)=\frac{(1-D_b)V_{dc}(k-2)I_a(k-1)}{2V_{dc}(k-2)-V_{dc}(k-3)}
\end{equation}

\begin{figure}[ht!]
\centering
    \includegraphics[width = 3 in, height = 1.6 in] {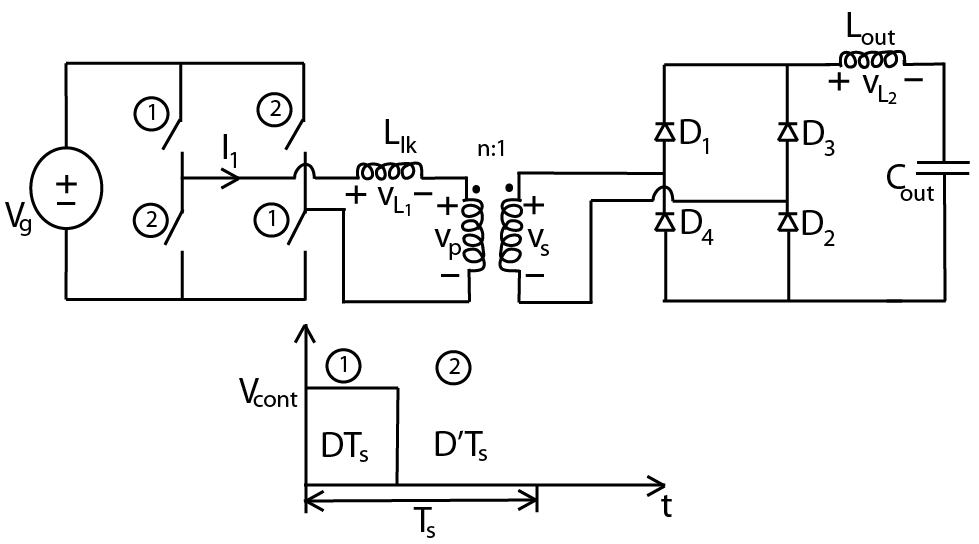}
    \caption{Circuit diagram of the PMU Block and the control switch pattern}
    \label{Circuit diagram of the PMU Block}
\end{figure}
\begin{figure}[ht!]
\centering
    \includegraphics[width = 3 in, height = 1 in] {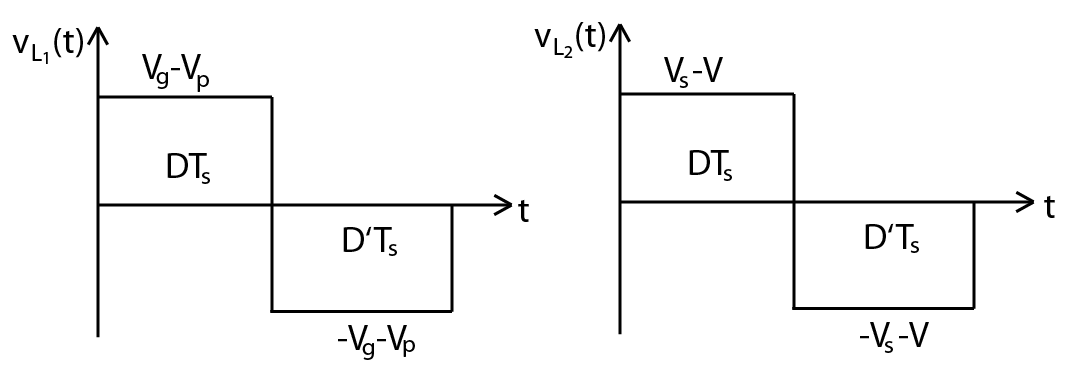}
    \caption{Inductive voltage pattern considering ideal active and passive switches  }
    \label{CSemiconductor Losses at switch 1 and 2}
\end{figure}
\begin{figure}[ht!]
\centering
    \includegraphics[width = 3 in, height = 2.8 in] {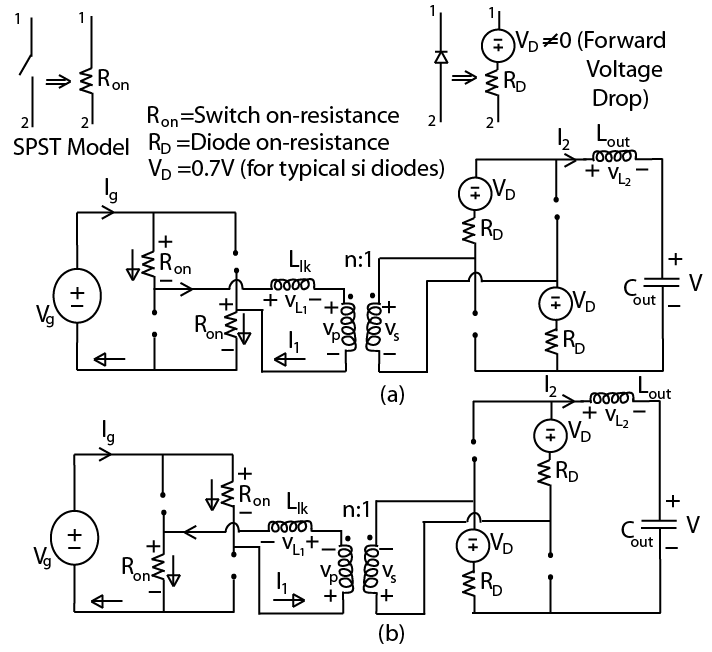}
    \caption{PMU circuit operation analysis considering non-ideal active and passive switches; (a) PMU in switch position-1 and (b) PMU in switch position-2}
\end{figure}
\begin{figure}[ht!]
\centering
    \includegraphics[width = 3 in, height = 1 in] {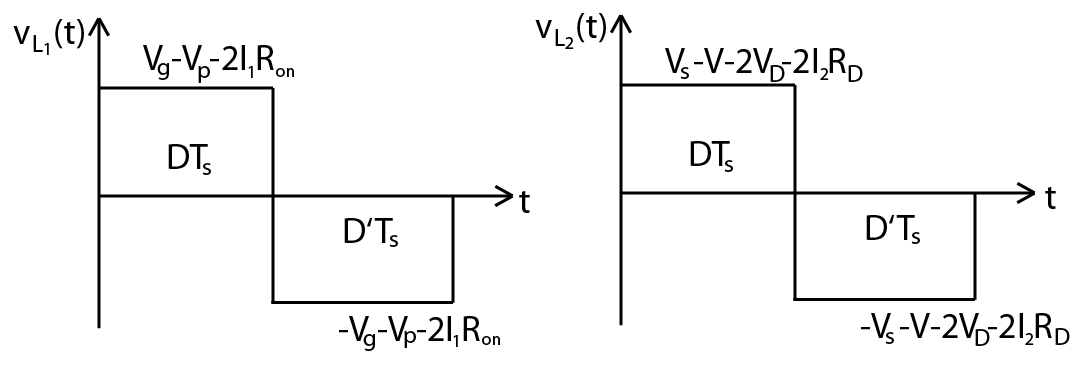}
    \caption{Inductive voltage pattern considering non-ideal active and passive Switches }
\end{figure}

\subsection{PMU Circuit Operation Analysis}
\subsubsection{Ideal Active and Passive Switches}
The MOSFETs in the PMUs are the active switches and each has an internal diode paralleled with a RC snubber circuit. When a gate signal is applied to the MOSFET it acts as the switch on-resistance and when the gate signal is turned off the current moves through the anti-parallel diode. $V_g$ is the input voltage for the PMUs and the control switch are delineated as 1 and 2 inside a circle in \ref{Circuit diagram of the PMU Block}. A linear transformer is used to lower the voltage into desired level as well as provide electrical isolation. The four diodes $D_1$,$D_2$,$D_3$ and $D_4$ are the passive switches. The $L_{lk}$ models the leakage inductance and has a damping effect on the circuit. The change in current through the output inductor $L_{out}$, and the change in voltage across the output capacitor $C_{out}$ are small compared through their nominal values over a switching period \cite{Vlatkovic_1992}-\cite{Madduri_2015}. Considering the control switch pattern in Fig. \ref{Circuit diagram of the PMU Block}, $D+D\textprime=1$ at switch position 1 during $DT_s$ subinterval
\begin{equation}\label{eq-1}
   V_{L_1}=V_g-V_p\hspace{5pt} and\hspace{5pt} V_{L_2}=V_s-V
\end{equation}
at switch position 2 during $D\textprime T_s$ subinterval
\begin{equation}\label{eq-2}
   V_{L_1}=-V_g-V_p\hspace{5pt} and\hspace{5pt} V_{L_2}=-V_s-V
\end{equation}
\begin{figure}[ht!]
\centering
    \includegraphics[width = 2.5 in, height = 1.2 in] {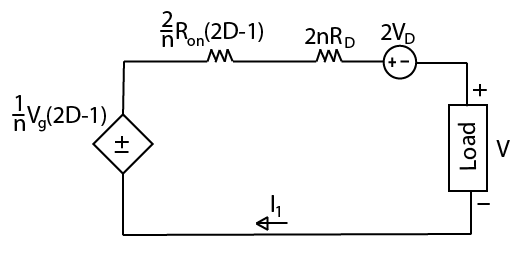}
    \caption{Input-to-output port of the PMU circuit acting in non-ideal switching mode}
\end{figure}


\begin{figure*}[ht!]
\centering
    \includegraphics[width = 6.5 in, height = 3 in] {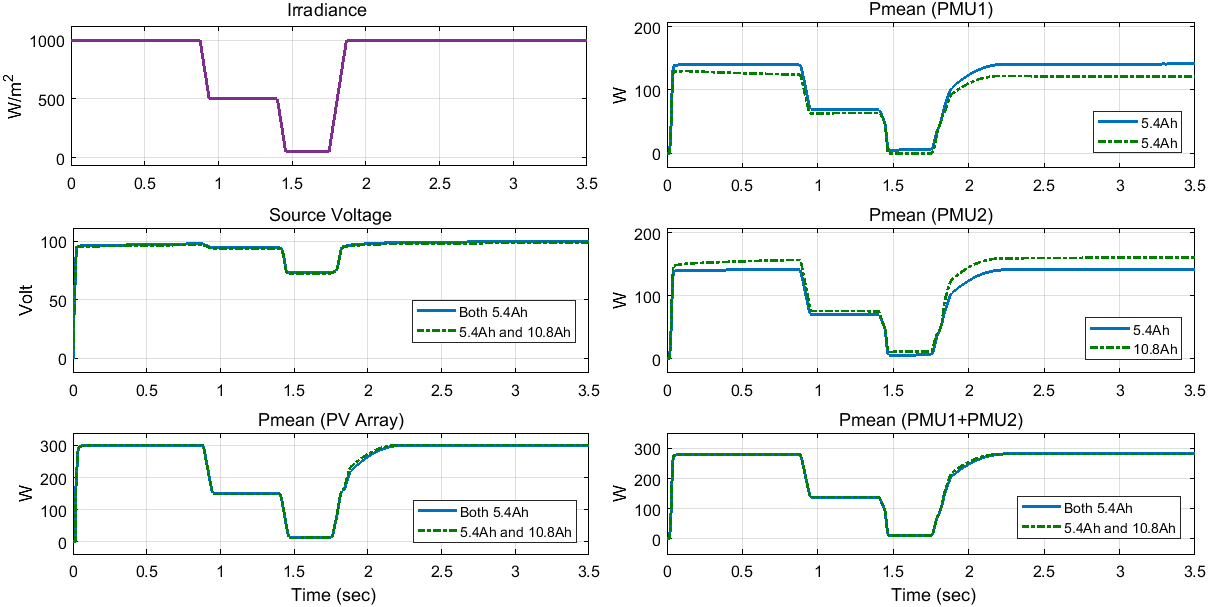}
    \caption{Simulation results for $12V$ batteries with same and different capacity in two separate PMUs in $50 kHz$ switching frequency for the active switches. The irradiance pattern was kept identical for each case. The green dotted line represents the case when the two separate PMUs have same battery capacity while the blue solid line depicts the case with two different battery capacity in two separate PMUs.}
    \label{100KHz_12V_Combined}
\end{figure*}
\begin{figure*}[ht!]
\centering
    \includegraphics[width = 6.5 in, height = 3 in] {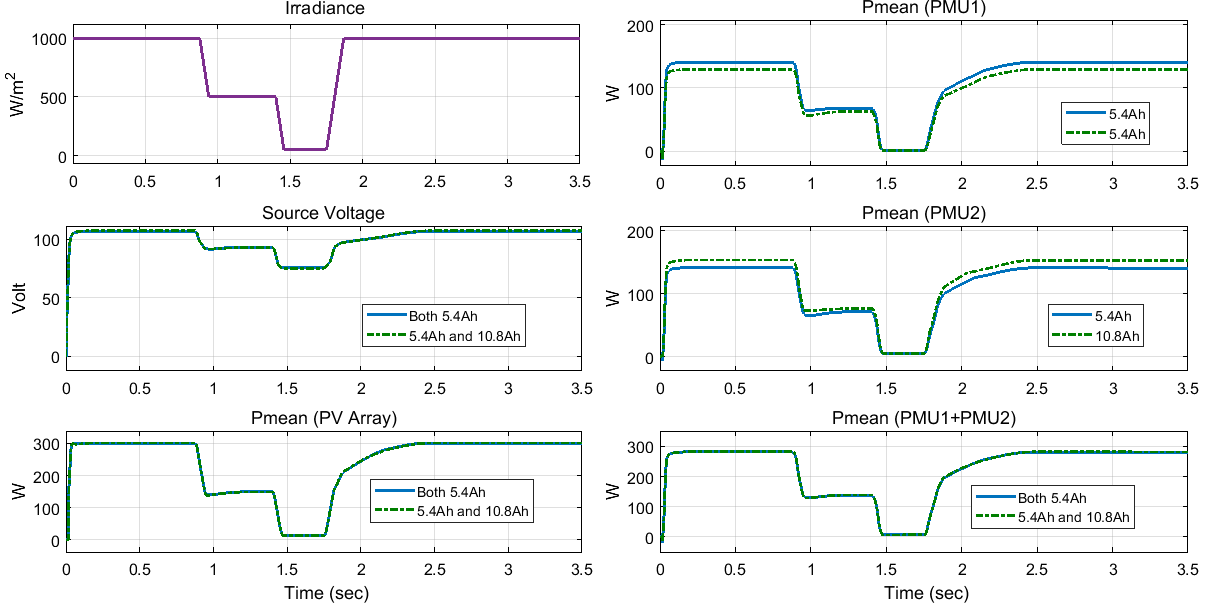}
    \caption{Simulation results for $36V$ batteries with same and different capacity in two separate PMUs in $50 kHz$ switching frequency for the active switches. The irradiance pattern was kept identical for each case. The green dotted line represents the case when the two separate PMUs have same battery capacity while the blue solid line depicts the case with two different battery capacity in two separate PMUs.}
    \label{50KHz_36V_Combined}
\end{figure*}
in both cases, small ripple approximation is considered here,
\begin{equation}\label{eq-3}
   V_p=nV_s
\end{equation}
Using volt-sec balance for inductor $L_{lk}$ and $L_{out}$ taking equation \eqref{eq-1}, \eqref{eq-2} and \eqref{eq-3}
\[(V_g-nV_s)D+(-V_g-nV_s)D\textprime=0\]
\begin{equation}\label{eq-4}
V_s=\frac{1}{n}V_g(2D-1)
\end{equation}
\[(V_s-V)D+(-V_s-V)D\textprime=0\]
\[V_s(2D-1)=V\]
from \eqref{eq-4}, voltage gain,
\begin{equation}\label{eq-5}
M(D)=\frac{V}{V_g}=\frac{1}{n}(2D-1)^2
\end{equation}
\subsubsection{Active and Passive Switches with Semiconductor Losses}
At switch 1
\begin{equation}\label{eq-6}
V_{L_1}=V_g-V_p-2I_1R_{on}
\end{equation}
\begin{equation}\label{eq-7}
V_{L_2}=V_s-V-V_D-2I_2R_{on}
\end{equation}
At switch 2
\begin{equation}\label{eq-8}
V_{L_1}=-V_g-V_p-2I_1R_{on}
\end{equation}
\begin{equation}\label{eq-9}
V_{L_2}=-V_s-V-2V_D-2I_2R_{on}
\end{equation}
Now using volt-second balance for $L_1$ and $L_2$ and $V_p=nV_S$ and $I_2=I_1$
\[(V_g-V_p-2I_1R_{on})D+(-V_g-V_p-2I_1R_{on})D\textprime=0\]
\[V_g(2D-1)-2I_1R_{on}=nV_s\]
\begin{equation}\label{eq-10}
V_s=\frac{1}{n}[V_g(2D-1)-2I_1R_{on}]
\end{equation}
and
\[(V_s-V-2V_D-2I_2R_D)D+(-V_s-V-2V_D-2I_2R_D)D\textprime=0\]
\[V_s(2D-1)=V+2V_D+2I_2R_D\]
\[\frac{1}{n}[V_g(2D-1)-2I_1R_{on}](2D-1)=V+2V_D+2I_2R_D\]
\begin{equation}\label{eq-11}
\frac{1}{n}V_g(2D-1)^2=\frac{2}{n}I_1R_{on}(2D-1)+V+2V_D+2nI_1R_D
\end{equation}

From \eqref{eq-11}
\[\frac{1}{n}V_g(2D-1)^2[1-\frac{2I_1R_{on}}{V_g(2D-1)}-\frac{2nV_D}{V_g(2D-1)^2}-\frac{2n^2I_1R_D}{V_g(2D-1)^2}]=V\]
Now, voltage gain,
\begin{dmath}\label{eq-12}
M\textprime(D)=\frac{V}{V_g}=M(D)[1-\frac{2I_1R_{on}}{V_g(2D-1)}-\frac{2nV_D}{V_g(2D-1)^2}-\frac{2n^2I_1R_D}{V_g(2D-1)^2}]
\end{dmath}

\section{Results and Analysis}

\begin{table*}[]
\centering
\caption{Performance Analysis of the Microgrid Architecture for Different Switching Frequencies and Battery Ratings}
\label{simulation_data_table}
\begin{tabular}{|l|l||l|l|l|l|l|l|l|l|l|l|l|l|l|l|}
\hline
Switching & Battery & Irradiance & \multicolumn{2}{l|}{PV Array} & 
\multicolumn{3}{l|}{Source Converter} & MPPT & \multicolumn{3}{l|}{PMU1} & \multicolumn{3}{l|}{PMU2} & Efficiency\\ 
\cline{4-5}\cline{6-8}\cline{10-12}\cline{13-15}
Frequency   &           & (W/m^2)    & V        & I                   & V      & I     & P_{mean}    & Duty   & V      & I     & P_{mean} & V      & I     & P_{mean}  & (Percent)  \\
            &           &            &          &                     &        &       &             & Cycle  &        &       &          &        &       &           &             \\\hline\hline
100 kHz     & 12V 5.4Ah     & 1000   & 51.88    & 5.798               & 99.09  & 3.033 & 300.8       & 0.4768 & 13.02  & 10.81 & 140.7    & 13.02  & 10.81 & 140.7     & 93.55 \\\cline{2-16}
            & 12V 5.4Ah     & 1000   & 51.88    & 5.798               & 98.28  & 3.059 & 300.8       & 0.4724 & 12.96  & 9.485 & 122.9    & 12.75  & 12.40 & 158.1     & 93.41 \\
            & 12V 10.8 Ah   &        &          &                     &        &       &             &        &        &       &          &        &       &           &       \\\cline{2-16}
            & 36V 5.4Ah     & 1000   & 51.87    & 5.799               & 115.6  & 2.600 & 300.8       & 0.5516 & 37.61  & 3.757 & 141.3    & 37.61  & 3.757 & 141.3     & 93.95 \\\cline{2-16}
            & 36V 5.4Ah     & 1000   & 51.88    & 5.798               & 114.9  & 2.616 & 300.8       & 0.5486 & 37.57  & 3.511 & 131.9    & 37.12  & 4.056 & 150.6     & 93.92 \\
            & 36V 10.8 Ah   &        &          &                     &        &       &             &        &        &       &          &        &       &           &           \\\hline\hline
 50 kHz     & 12V 5.4Ah     & 1000   & 51.89    & 5.797               & 99.94  & 3.007 & 300.8       & 0.4810 & 13.02  & 10.83 & 141.0    & 13.02  & 10.83 & 141.0     & 93.55 \\\cline{2-16}
            & 12V 5.4Ah     & 1000   & 51.92    & 5.793               & 98.82  & 3.042 & 300.8       & 0.4749 & 12.95  & 9.363 & 121.3    & 12.75  & 12.56 & 160.2     & 93.41 \\
            & 12V 10.8 Ah   &        &          &                     &        &       &             &        &        &       &          &        &       &           &       \\\cline{2-16}
            & 36V 5.4Ah     & 1000   & 51.88    & 5.797               & 106.3  & 2.827 & 300.8       & 0.5123 & 37.65  & 3.721 & 140.1    & 37.65  & 3.721 & 140.1     & 93.95 \\\cline{2-16}
            & 36V 5.4Ah     & 1000   & 51.92    & 5.793               & 106.4  & 2.824 & 300.8       & 0.5122 & 37.65  & 3.728 & 140.4    & 37.12  & 3.780 & 140.4     & 93.92 \\
            & 36V 10.8 Ah   &        &          &                     &        &       &             &        &        &       &          &        &       &           &           \\\hline
\end{tabular}
\end{table*}

The DC microgrid structure is simulated using MATLAB 2017a Simulink. SunPower SPR-315E-WHT-D Module which follows NREL (National Renewable Energy Laboratory) System Advisor Model (Jan. 2014), is used as the PV module which takes two inputs of Sun irradiance and Cell temperature. At $1000 W/m^2$ and $40$ deg. Celsius the maximum power point is $300.8 W$. Inside the source converter block, a MPPT Control and a Boost average converter are employed to achieve the maximum power point as well as boosting the voltage before transmission. 
The MPPT control uses Perturb and Observation (P\&O) algorithm which takes instantaneous voltage and current of the solar module as the input. There is a enable pin in the MPPT block which is switched to 1 to allow the continuous opetation of the algorithm. Four other control parameters for P\&O algorithm are Initial value for $D_b$ output ($D_{b_{init}}$), Upper limit for $D_b$ ($D_{b_{max}}$), Lower limit for $D_b$ ($D_{b_{min}}$), Increment value used to increase/decrease ($\Delta D_b$) where $D_b$ is the Boost converter duty cycle.The boost converter uses the output duty cycle of the PO and generates the desired voltage realizing \eqref{eq1-} and \eqref{eq2-}. A constant of $10^{-6}$ is used to avoid the divide by zero condition in \eqref{eq2-}. Inside the PMUs, the value of leakage inductance $L_{lk}$, outer inductance $L_{out}$ and the outer capacitance is $26\mu H$, $10\mu H$ and $7500 \mu F$.
The nominal power of the linear transformer is set to 150 VA  in case of equal loads and frequency is adjusted according to the switching frequency of the active switches. The MOSFETs' on-resistance $R_{on}$ and diode resistance $R_D$  has the value $0.1 \Omega$ and $0.01 \Omega$ respectively. Lead acid battery is used with nominal voltage of $12V$, rated capacity $5.4Ah$, initial state of charge $75\%$ and battery response time is $50min$. The irradiation was varied from $1000 W/m^2$ to $500 W/m^2$, $500 W/m^2$ to $50W/m^2$ and then back to $1000 W/m^2$ from $50W/m^2$ for $12V$ and $36V$ batteries with same and different battery capacity in two different switching frequencies of $50 kHz$ and $100 kHz$. Fig. \ref{100KHz_12V_Combined} shows the simulation results for $12V$ batteries with same and different capacity in two separate PMUs in $100 kHz$ switching frequency for the active switches. The output voltage of the source converter tends to show minor change due to fluctuation of the irradiance which is very encouraging. The average power is divided into PMUs equally when the battery parameters are uniform. The battery with higher rated capacity draws higher power resulting in unequal power sharing between the two PMUs although the source converter output voltage is almost the same as the former case. The average power from the PV array and the sum of the average power of the PMUs are very closely matched in both cases and proportionate to the irradiance pattern. An interesting observation is that when the irradiance jumps back to $1000 W/m^2$ from $50W/m^2$ the average power slows down to move to $300.8$ towards the end.

Table \ref{simulation_data_table} shows the simulation results for two different battery voltage in two disparate switching frequency. For $36V$ batteries the transformer ratings were adjusted to maintain the output of the PMUs to the desired level. The MPPT duty cycle moved to a different point resulting in a slightly different source converter output voltage around $110V$ but keeps average power to $300.8W$. The efficiency estimated in all different cases is more than $93\%$ which ensures the stability of the system in different battery voltage level and frequency.

\section{Conclusion}

The design presented in this paper is a simplified version of the scalable DC microgrid architecture. The control phenomenon of the load converter and the load side of the home PMU configuration are different from those of the generic structure proposed in \cite{Madduri_2015}. Basic PMW signals are used instead of phase-shifted PWM as the switch control attributes in the PMUs. The look-up table generated from multiple simulations is an attempt to establish a one-way communication method for smart load sharing among different PMUs. One-way communication will reduce the complexities of the electrical circuits and can prove to be a cost effective solution which can have a major effect in popularizing the DC microgrid in the developing countries. Moreover, phase-shifted PWM can be implemented for zero voltage switching of the converter and practical implication of the system.






\begin{thebibliography}{1}


\bibitem{IEA_2011}
''WORLD ENERGY OUTLOOK 2011 FACTSHEET'', \emph{IEA World Energy Outlook
2011}, November 2011.

\bibitem{Alstone_2015}
P. Alstone, D. Gershenson, and D. M. Kammen, ``Decentralized energy systems for clean
electricity access'', \emph{Nature Climate Change}, 2015.


\bibitem{Komatsu_2011}
S. Komatsu, S. Kaneko, and P. P. Ghosh, ``Are micro-benefits negligible? The implications
of the rapid expansion of Solar Home Systems (SHS) in rural Bangladesh for
sustainable development'', \emph{Energy Policy}, 2011.

\bibitem{Bardouille_2012}
P. Bardouille, P. Avato, J. Levin, A. Pantelias, and H. Engelmann-Pilger, ``From gap
to opportunity: Business models for scaling up energy access'', \emph{International Finance
Corporation}, 2012

\bibitem{Boroyevich_2010}
Dushan Boroyevich, Igor Cvetkovic, Dong Dong, et al, ``Future electronic power
distribution systems a contemplative view'' \emph{12th International Conference
on Optimization of Electrical and Electronic Equipment}, pages 1369–1380. IEEE,
May 2010.

\bibitem{Quetchenbach_2013}
T. G. Quetchenbach, M. J. Harper, J. Robinson IV, et al., ``The GridShare solution: a
smart grid approach to improve service provision on a renewable energy mini-grid in
Bhutan.'',
\emph{Environmental Research Letters},8(1):014018, 2013.

\bibitem{Vlatkovic_1992}
V. Vlatkovic, J. A. Sabate, R. B. Ridley, F. C. Lee, and
B. H. Cho, ``Small-Signal Analysis of the Phase-Shifted PWM
Converter'',
\emph{IEEE Transactions on Power Electronics},7(1),pages 128-135, January 1992.

\bibitem{Schutten_2003}
M. J. Schutten and D. A. Torrey, ``Improved small-signal analysis for
the phase-shifted PWM power converter'',
\emph{IEEE Transactions on Power Electronics},vol. 18, no. 2, pp. 659–669, Mar. 2003.

\bibitem{Madduri_2015}
P. A. Madduri, J. Poon, J. Rosa, M. Podolsky, E. Brewer, S. Sanders, ``A scalable dc microgrid architecture for rural electrification in emerging regions'',
\emph{Applied Power Electronics Conference and Exposition},7(1),pages 703-708, Mar 2015.

\bibitem{Stevens_1996}
W. Stevens and G. P. Corey, ``A study of lead-acid battery efficiency
near top-of-charge and the impact on pv system design'',
\emph{Photovoltaic
Specialists Conference, 1996.}, Conference Record of the Twenty Fifth
IEEE, pp. 1485–1488, 1996.


\bibitem{Brent_2010}
A. C. Brent and D. E. Rogers, ``Renewable rural electrification: Sustainability
assessment of mini-hybrid off-grid technological systems in the
African context'',
\emph{Renewable Energy}, vol. 35, no. 1, pp. 257–265, Jan.
2010.

\end{thebibliography}
%

\end{document}